\documentclass[preprint,aps]{revtex4}
\pdfoutput=1
\usepackage{dcolumn}
\usepackage{bm}
\usepackage{graphicx}
\usepackage{epsfig}
\usepackage{amsmath}
\usepackage{amssymb}
\usepackage{mathrsfs}
\usepackage{ulem}
\usepackage{color}
\usepackage{hyperref}
\usepackage{float}
\allowdisplaybreaks

\def\fun#1#2{\lower3.6pt\vbox{\baselineskip0pt\lineskip.9pt
  \ialign{$\mathsurround=0pt#1\hfil##\hfil$\crcr#2\crcr\sim\crcr}}}

\def\simlt{\stackrel{<}{{}_\sim}}
\def\simgt{\stackrel{>}{{}_\sim}}

\input epsf

\newcommand{\be}{\begin{equation}}
\newcommand{\ee}{\end{equation}}
\newcommand{\bea}{\begin{eqnarray}}
\newcommand{\eea}{\end{eqnarray}}

\begin{document}

\begin{flushright}
\vspace{-0.2cm}EFI-15-6\\
\end{flushright}


\title{ CP-odd component of the lightest neutral Higgs boson \\ in the MSSM}

\vspace*{0.2cm}

\author{
\vspace{0.2cm}
\mbox{\bf Bing Li$^{a}$ and Carlos E.~M.~Wagner$^{a,b,c}$}
 }
\affiliation{
\vspace*{.5cm}
$^a$  \mbox{Enrico Fermi Institute, University of Chicago, Chicago, IL 60637}\\
$^b$  \mbox{Kavli Institute for Cosmological Physics, University of Chicago, Chicago, IL 60637}\\
$^c$ \mbox{High Energy Physics Division, Argonne National Laboratory, Argonne, IL 60439}\\
}

\begin{abstract}
The Higgs sector of the Minimal Supersymmetric Extension of the Standard Model may be described with
a two Higgs doublet model with properties that depend on the soft supersymmetry breaking parameters.
For instance, flavor independent CP-violating phases associated with the gaugino
masses, the squark trilinear mass parameters and the Higgsino mass parameter $\mu$ may lead
to sizable CP-violation in the Higgs sector.
For these CP-violating effects to affect the properties of the recently observed SM-like Higgs resonance, the non-standard
charged and neutral Higgs bosons masses must be of the order of the weak scale, and both $\mu$
as well as the trilinear stop mass parameter $A_t$ must be of the order or larger than the
stop mass parameters.  Constraints on this possibility come from direct searches for non-standard
Higgs bosons, precision measurements on the lightest neutral Higgs properties, including its mass,
and electric dipole moments. In this article, we discuss these constraints within the MSSM, trying to evaluate the possible
size of the CP-odd component of the lightest neutral Higgs boson, and the possible experimental tests
of this CP-violating effect at the LHC.
\end{abstract}
\thispagestyle{empty}

\maketitle

 \section{Introduction}

 The Minimal Supersymmetric Extension of the Standard Model (MSSM) is an attractive scenario that leads to a well defined spectrum of particles
 at low energies, with dimensionless couplings that are related to the Standard Model (SM) ones by symmetry relations. For third generation superpartners with masses of
 the order of the TeV scale, this scenario leads to radiative electroweak symmetry breaking, it is consistent with unification of couplings at high energies~\cite{reviews}
 and in the presence of R-Parity contains a Dark Matter particle identified with the lightest neutralino~\cite{Goldberg:1983nd},\cite{Ellis:1983ew}.

 The Higgs sector of the theory contains two doublets, and at tree-level supersymmetry demands it to be of type-II and CP-conserving, with an
 upper bound on the lightest CP-even Higgs mass equal to the gauge boson mass $M_Z$.  These properties are modified at the quantum
 level~\cite{Haber:1990aw}--\cite{FeynHiggs}. On one hand, as it is well known, in the absence of CP-violation, the upper bound on the lightest CP-even Higgs mass is no longer $M_Z$ but could be raised to values of order 130~GeV for stop masses of the order of a few TeV and sizable values of the trilinear stop mass parameter $A_t$.
 The observed values of the Higgs mass may be then well explained in this scenario~\cite{Heinemeyer:2011aa}. On the other hand, radiative
 corrections also induce deviations from the type-II behavior that become more prominent for large values of the ratio of vacuum expectation
 values $\tan\beta$ and small values of the non-standard Higgs boson masses.

 CP violation in the effective two Higgs Doublet Model (2HDM) can be induced by phases of the soft SUSY-breaking parameters at the loop level ~\cite{Pilaftsis:1998dd}--\cite{Lee:2012wa}. In this model, the lightest neutral Higgs is no longer a CP-eigenstate, but a mixture of CP-even and CP-odd states. The presence of CP-violation in the mass parameters of the theory is natural within the MSSM, and may be related to the mechanism that explains the baryon asymmetry in the universe~\cite{Shu:2013uua}. Indeed, it is known that the CP-violation present in the SM is not sufficient to explain the baryon asymmetry and new CP-violating effects are necessary.
The presence of CP-violation in the Higgs sector may lead to a modification of the neutral Higgs properties that may be tested at the LHC in the near future.
In particular, the recently discovered Higgs boson at the LHC~\cite{ATLASCMS} may be the lightest of the three neutral states, with a non-vanishing CP-odd component.

Due to the current lack of observation of CP-violation observables beyond those present in the Standard Model, in particular the electron and the neutron electric dipole moments~\cite{Baker:2006ts}--\cite{Wang:2014hba},  large phases in the gaugino  mass and  the $\mu$
parameters tend to be  in conflict with a light supersymmetric spectrum~\cite{Brhlik:1998zn}--\cite{Ibrahim:2007fb}.  These restrictions may be alleviated by assuming large values of the first and
second generation slepton and squark masses. Even in this case, two-loop CP-violating effects may be large enough to lead to observable CP-violating
effects which may be in conflict with present experimental bounds.

In a recent article~\cite{Arbey:2014msa}, the authors analyzed the  CP-odd mixing of the heavy neutral states, allowed by the
current flavor physics, Higgs and electric dipole moment constraints.  In this article, we shall concentrate on an analysis of the CP-odd component
of the lightest neutral Higgs in the MSSM,
given all available constraints from both the experimental and the theoretical side (for a previous study, see Ref.~\cite{Chakraborty:2013si}). We provide an analytical understanding of the parameters that control
this CP-odd component and analyze the impact of these parameters on the Higgs observables. We shall compare these analytical results with the
ones provided by  CPsuperH2.3, which is used to calculate the masses of neutral Higgs, their production rates, decay widths and couplings with other particles~\cite{Lee:2003nta,Lee:2007gn,Lee:2012wa}.  Based on this analysis, we found that if the stop particles are assumed to be lighter than a few TeV, the requirement of obtaining a 125.5~GeV Higgs mass already puts a strong constraint to the parameter space and already restricts  the possibility of a CP-odd mixing higher than
about 10\%.  Moreover, the current measurements of the lightest CP-even Higgs production rates puts further constraints on this possibility and so does the non-observation of the electron, neutron
and Mercury electric dipole moments. Based on these facts, we study the capability of the
LHC to detect the small CP-odd components of the lightest neutral Higgs within the MSSM.

This article is organized as follows. In section~\ref{sec:CPodd} we describe the relevant parameters controlling the CP-violating effects in the neutral Higgs sector. In section~\ref{sec:Mass} we provide analytical formulae for the neutral Higgs mass matrix elements and describe the interrelation between the CP-odd component of the lightest Higgs and its mass. In section~\ref{sec:decay}  we describe similar constraints affecting the decay branching ratios of the lightest neutral Higgs boson. In sections~\ref{sec:edm} and~\ref{sec:flavor} we discuss the constraints coming from electric dipole moments and flavor physics. We discuss the possible measurement of the lightest neutral Higgs CP-odd component at the LHC in section~\ref{sec:LHC}. We reserve
section~\ref{sec:conclusions} for our conclusions.

 \section{CP-odd Component of the Lightest Neutral Higgs Boson}
\label{sec:CPodd}

The CP-violating phases in the low energy 2HDM may come in the MSSM soft breaking parameters. Since these CP-violating effects
are induced at the loop-level, the only relevant phases are the ones associated with supersymmetric  particles that couple strongly to the Higgs bosons, namely the stops, sbottoms and staus, and the gluinos that couple strongly to these particles ~\cite{Pilaftsis:1998dd}--\cite{Lee:2012wa}.  The relevant complex phases are then the ones of the trilinear soft couplings of the stops, sbottoms  and staus to the Higgs field, $\Phi_{A_{t}}$, $\Phi_{A_{b}}$, $\Phi_{A_{\tau}}$, respectively, the phase of the gluino mass parameter $ \Phi_{M_{\tilde{g}}}$,  and the one of the Higgsino mass parameter $\mu$, $\Phi_{\mu}$.   Besides, one should also consider the variations of the magnitude of $\tan{\beta}$, $|A_{t,b,\tau}|$, $|M_{\tilde{g}}|$, $|\mu|$, $m_{H^+}$, and the mass parameter $M_{\rm SUSY}$, that controls the overall third generation mass scale.  CP-violating effects are induced by non-decoupling threshold corrections and become relevant whenever the imaginary part of $\mu A_{t,b,\tau}$ and/or of $\mu M_{\tilde{g}}$ is non-zero and of the order or larger than the square of the third generation sfermion masses, which we shall assume to be of the order of a few TeV.

Our objective is to study regions of parameter space in which a large CP-odd component of the lightest neutral Higgs is present.  Since this component may only be induced by mixing between the would-be CP-even and CP-odd Higgs states,  it is clear that the heavier neutral Higgs bosons should be light, with masses not much larger than the weak scale.  Such values of the non-standard Higgs boson masses lead naturally to large variations of the fermion couplings to the lightest neutral Higgs with respect to the Standard Model ones, and  also leads to a reduction of the lightest neutral Higgs mass via the mixing with the other neutral states.

In the analysis of the parameters of the model, we shall require the mass of the lightest neutral state to be consistent with the  measured value of about 125.5~GeV. Due to theoretical uncertainties in the calculation of the neutral Higgs masses, which is of the order of 3~GeV, we shall retain values of the parameters which lead to a Higgs mass between 122.5 and 128.5~GeV.  Moreover, the bottom and tau couplings of the lightest Higgs boson cannot differ significantly from the ones of the SM without leading to significant variations of the Higgs decay branching ratios, in conflict with observations at the ATLAS and CMS experiments. In general, since the electroweak gauge boson couplings of the lightest Higgs tend to be close to the SM ones, variations of the effective bottom coupling $g_{H_1 b\bar{b}}$ of more than about 20\% with respect to the SM (leading to variations of the branching ratio of the decay of the Higgs boson to pairs of gauge bosons of about 30\%) are disfavored by data.

Although currently only one Higgs boson has been detected, there is information on the possible presence of additional Higgs bosons within the MSSM due to the non-observation of non-standard Higgs signatures. Currently, the strongest bounds on the presence of non-standard neutral Higgs bosons come from the searches of the gluon fusion or $bb\Phi$ production of heavy neutral Higgs bosons at the LHC, with subsequent decays into tau pairs~\cite{CMS:2011axa},\cite{Khachatryan:2014wca},\cite{Aad:2014vgg}.  These searches become particularly efficient for large values of $\tan\beta$ and low values of the charged Higgs mass $m_{H^+}$, for which the production rate is large.  These searches, combined with previous LEP results, gave a strong constraint on the $\tan\beta-M_{A}$ two dimensional plane (CP-violation was not considered in the LHC analyses). A small window of $\tan\beta$ survives in lower-$M_{A}$ region, where larger CP-violation is most likely to arise. In particular, for non-standard Higgs boson masses of the order of the weak scale, values of $\tan\beta > 10$ are strongly restricted by the searches performed by the CMS and ATLAS experiments.

\section{Constraints on CP violation in the Higgs sector from the lightest neutral Higgs mass}
\label{sec:Mass}

Since the LHC has measured a Higgs  boson with  mass  around 125.5~GeV, it is natural to identify it with the lightest neutral Higgs boson, which tends to have SM-like properties when masses of the heavier Higgs bosons are larger than 200~GeV. For stop masses of the order of a few TeV, this strongly restricts the plausible MSSM parameter space. As the charged Higgs mass goes up, the lightest CP-even Higgs mass depends mostly on $|X_t|$, with $X_t  = A_t - \mu^*/\tan\beta$~\cite{Pilaftsis:1998dd}--\cite{FeynHiggsCP}.  For values of the stop masses of the order of a few TeV a maximum value of the order of 130~GeV is obtained for values of $|X_t|$ of about 2.4~$M_{\rm SUSY}$, for large values of the charged Higgs mass, and goes smoothly down for smaller values
of $m_{H^+}$. Thus acceptable values of the Higgs mass are obtained for values of $|X_t|$ larger than $M_{\rm SUSY}$ but not larger than $3 M_{\rm SUSY}$.
For values of $|X_t|$ larger than 3~$M_{\rm SUSY}$ the lightest CP-even Higgs mass decreases sharply and, in addition, problems with vacuum stability
may be generated~\cite{VacStab}.

To explore the correlation between the CP-odd component and the mass of the lightest Higgs, we'll start from the $3\times3$ mass matrix, defining the mixing  between the would-be CP-even components of the two Higgs doublets and the CP-odd Higgs boson in the absence of CP-violating effects, $\phi_1$, $\phi_2$ and $a$, respectively. Let's separate out the tree-level terms and investigate the contributions from the CP-violating phases, taken as small perturbations here, to see how those perturbations affect the mass eigenstates of the neutral Higgs sector. The full mass matrix can be written as,
\begin{eqnarray}
  M^2 & = & M_{Tree}^2+M_{Loop}^2 \\
  & = &
  \left(
  \begin{array}{ccc}
  M_{a}^2 s_{\beta}^2+M_{z}^2 c_{\beta}^2 & -(M_{a}^2+M_{z}^2)s_{\beta}c_{\beta} & 0\\
   -(M_{a}^2+M_{z}^2)s_{\beta}c_{\beta} & M_{a}^2 c_{\beta}^2+M_{z}^2 s_{\beta}^2 & 0 \\
  0 & 0 & M_a^2
  \end{array}
  \right)
  +
  \left(
  \begin{array}{ccc}
    \Delta_{11} & \Delta_{12} & \delta_1 \\
   \Delta_{21} & \Delta_{22} & \delta_2 \\
  \delta_{1} & \delta_{2} & 0
  \end{array}
  \right)
\end{eqnarray}
where $\delta_{i}$,$\Delta_{ij}$ can be considered as perturbations and we'll investigate their effects on Higgs mass in the following. With the relative phase $\xi$ between the two Higgs doublets set to be zero, $\delta_{i}$, $\Delta_{ij}$ can be expanded as follows,
\begin{equation}
\begin{split}
  \delta_{1} &=v^2(Im(\lambda_{5})s_{\beta}+Im(\lambda_{6})c_{\beta})\\
  \delta_{2} &=v^2(Im(\lambda_{5})c_{\beta}+Im(\lambda_{7})s_{\beta})\\
  \Delta_{11} &=-v^2(2\lambda_1c_{\beta}^2+2Re(\lambda_5)s_{\beta}^2+2Re(\lambda_6)s_{\beta}c_{\beta})- M_{Z}^2c_{\beta}^2\\
  \Delta_{12} &=\Delta_{21} =-v^2(\lambda_{34}s_{\beta}c_{\beta}+Re(\lambda_6)c_{\beta}^2+Re(\lambda_7)s_{\beta}^2)+M_Z^2s_{\beta}c_{\beta}\\
  \Delta_{22} &=-v^2(2\lambda_2s_{\beta}^2+2Re(\lambda_5)c_{\beta}^2+2Re(\lambda_7)s_{\beta}c_{\beta})-M_{Z}^2s_{\beta}^2
\end{split}
\end{equation}

The values of the quartic couplings may be found in Ref.~\cite{Pilaftsis:1999qt}.
In order to understand the main effects, we should go to the Higgs basis ($\{\phi_1,\phi_2\}$$\to$$\{h_1,h_2\}$) by rotating by the angle $\beta$, which becomes the proper diagonalization angle in the decoupling limit. The transformation matrix O links the 3 neutral Higgs further with their mass eigenstates by $\{h_1, h_2, a\}^{T}=O\{H_1, H_2, H_3\}^{T}$, thus $H_1$ can be expanded as $H_1=O_{11}h_1+O_{21}h_2+O_{31}a$. In this case, we  get,
\begin{equation}
 \label{}
 \begin{split}
  &O M^2_{\rm diag} O^{T}\\
  =&
  \left(
  \begin{array}{ccc}
 M_{Z}^2 \cos^2 2 \beta &  M_Z^2 \cos 2\beta \sin 2 \beta  & 0 \\
   M_Z^2 \cos2\beta \sin2\beta    & \left( m_{a}^2 + M_Z^2 \sin^22\beta \right) & 0 \\
  0 & 0  & m_{a}^2
  \end{array}
  \right)
  +
  \left(
  \begin{array}{ccc}
   c_{\beta} & s_{\beta}  & 0 \\
  -s_{\beta} & c_{\beta} & 0 \\
  0 & 0 & 1
  \end{array}
  \right)
  \left(
  \begin{array}{ccc}
   \Delta_{11} & \Delta_{12} & \delta_1 \\
   \Delta_{12} & \Delta_{22} & \delta_2 \\
   \delta_{1} & \delta_{2} & 0
  \end{array}
  \right)
  \left(
  \begin{array}{ccc}
   c_{\beta} & -s_{\beta} & 0 \\
   s_{\beta} & c_{\beta} & 0 \\
   0 & 0 & 1
  \end{array}
  \right) \\
  \nonumber
  \end{split}
\end{equation}

\noindent
\begin{equation}
 \begin{split}
  &=
  \left(
  \begin{array}{ccc}
   M_Z^2 \cos^22\beta +\eta & \theta  & \xi_2 \\
 \theta & m_{a}^2+\ M_Z^2 \sin^2 2 \beta + \rho  & \xi_1 \\
  \xi_2 & \xi_1 & m_a^2
  \end{array}
  \right)
 \end{split}
\end{equation}
where $M^2_{\rm diag}$ is the eigenvalue matrix and
\begin{equation}
\begin{split}
  \xi_1&= -\delta_1s_{\beta} + \delta_2c_{\beta}\\
  \xi_2&=\delta_1c_{\beta}+\delta_2s_{\beta}\\
  \theta &=  (\Delta_{22} - \Delta_{11}) \sin\beta \cos\beta +  \Delta_{12} \cos2\beta - M_Z^2 \cos 2\beta \sin 2 \beta \\
  \eta &= \Delta_{11} c^2_\beta + \Delta_{22} s^2_\beta + \Delta_{12} \sin2\beta
\end{split}
\end{equation}

In the result of equation(4), we can see that the final corrections to $m_{H_1}^2$ come from the three terms, $\xi_2, \theta, \eta$. In this limit, $\xi_2$ defines the strength of the mixing between a and h, i.e. it fixes the CP-odd component of the lightest Higgs.  Defining the parameter $Y_t = A_t + \mu^* \tan\beta$, one can demonstrate that, at one loop
\begin{equation}\
\label{eq:eta}
\eta = \frac{3 h_t^4 v^2 \sin^4\beta}{8 \pi^2} \left[ \log\left( \frac{M_{\rm SUSY}^2}{m_t^2} \right)+ \frac{|X_t|^2}{M_{\rm SUSY}^2} \left( 1 - \frac{|X_t|^2}{12 \ M_{\rm SUSY}^2} \right) \right]
\end{equation}
\begin{equation}
\label{eq:theta}
\begin{split}
\theta  = - M_Z^2 \cos 2\beta \sin 2\beta
 +  \frac{ 3 h_t^4 v^2 \sin^2\beta \sin2\beta}{16 \pi^2} \left[ \log\left( \frac{M_{\rm SUSY}^2}{m_t^2} \right) \right.\\
\left.+ \frac{ |X_t|^2}{2 M_{\rm SUSY}^2} + {\rm Re} \left( \frac{X_t Y_t^* }{2 M_{\rm SUSY}^2} \left(1  - \frac{|X_t|^2}{6 M_{\rm SUSY}^2} \right) \right)
\right]
\end{split}
\end{equation}
\begin{equation}
\label{eq:xi2}
\xi_2 =  {\rm Im} \left(
 \frac{ 3 h_t^4 v^2 \sin^2\beta \sin2\beta}{32 \pi^2} \left[
\frac{ X_t Y_t^*}{M_{\rm SUSY}^2}\left(1  - \frac{|X_t|^2}{6 M_{\rm SUSY}^2}  \right) \right] \right)
\end{equation}
~\\
where $v \simeq 246$~GeV is the Higgs vacuum expectation value.  The above equations provide a generalization of the expressions for the  Higgs
 mixing parameters in terms of $X_t$ and $Y_t$ in the CP-conserving case~\cite{Carena:2014nza}. The parameter $\eta$ displays the well known one-loop radiative
 corrections to the lightest (would be CP-even) Higgs mass, which
are maximized for values of the stop mixing parameter $|X_t| = \sqrt{6} \ M_{\rm SUSY}$.  Notoriously, for the same values of the stop mixing parameter
the parameter $\xi_2$ vanishes. Hence, a sizable CP-odd component of the lightest neutral Higgs boson is always associated with departures from
the maximal values of its mass.

\begin{figure}[H]
    \centering
    \includegraphics[width=9.cm]{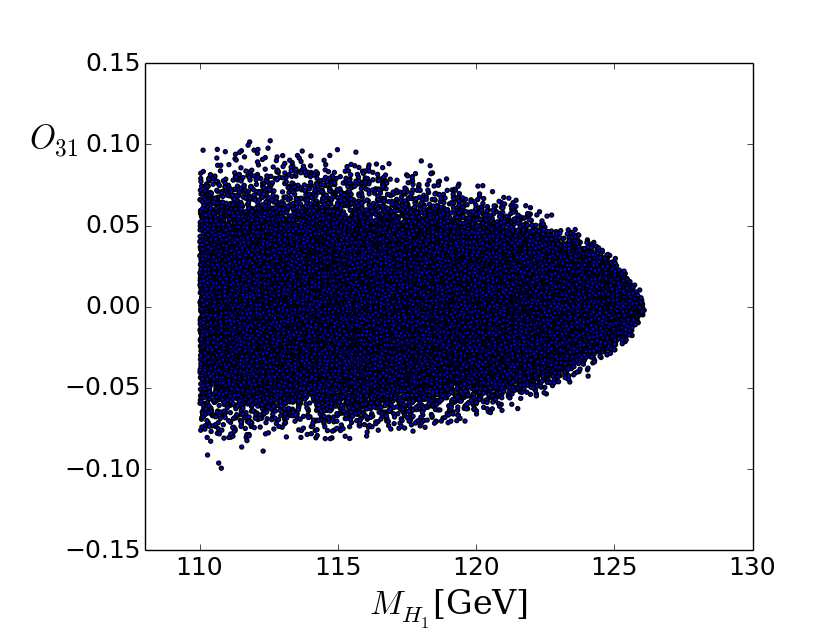}
    \caption{Correlation between the $H_1$ CP-odd component and its  mass for $\tan\beta=$~5.5 and a charged Higgs mass $M_{H^+}=260$~GeV.  The moduli and phases of all
    relevant parameters $A_f$, $M_{\tilde{g}}$ and $\mu$ were varied in the range explained in the text and the overall stop mass scale $M_{\rm SUSY}$ was fixed at 2~TeV.}
    \label{MHCPodd1}
\end{figure}

\begin{figure}[H]
    \centering
    \includegraphics[width=9.cm]{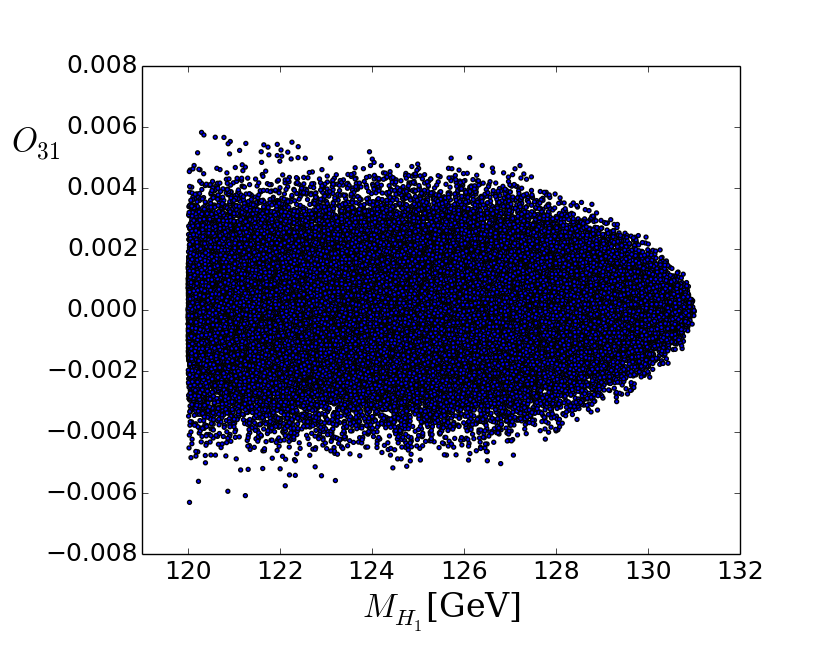}
    \caption{Correlation between the $H_1$ CP-odd component and its mass for  $\tan\beta=$~20 and a charged Higgs mass $M_{H^+}=800$~GeV. The moduli and phases of all
    relevant parameters $A_f$, $M_{\tilde{g}}$ and $\mu$ were varied in the range explained in the text and the overall stop mass scale $M_{\rm SUSY}$ was fixed at 2~TeV.}
\label{MHCPodd2}
\end{figure}

The above property is clearly shown in Figures~\ref{MHCPodd1} and~\ref{MHCPodd2} where we display the value of the CP-odd
component of the lightest neutral Higgs against its mass, obtained by the CPsuperH code~\cite{Lee:2007gn},\cite{Lee:2012wa}.
for two different values of $\tan\beta$ and the charged Higgs boson mass, consistent with
the current experimental bounds coming from direct searches for non-standard Higgs bosons at the LEP and LHC experiments. During this procedure, $400,000$ points were randomly generated and uniformly scattered all over the space spanned by the relevant parameters. We choose the values of the supersymmetry breaking parameter $M_{\rm SUSY} = 2$~TeV and the rest of the parameters were varied in the following ranges :  $A_t$ from 2~TeV to 6~TeV, $|\mu|$ from 2~TeV to 6~TeV, $\Phi_{M_3}$, $\Phi_{A}$, $\Phi_{\mu}$, $\Phi_{M_2}$ from $-180^{\circ}$ to $+180^{\circ}$, $|M_3|$ from 500~GeV to 3~TeV. The hierarchy factor $\rho$, denoting the difference between the masses of the first and second generation sfermions and the third generation ones plays only a small role in this analysis and was chosen to be equal to one. From this plot we see that there is an upper limit for the lightest neutral Higgs mass around 127~GeV for a charged Higgs mass, $M_{H^+} = 260$~GeV and $\tan\beta = 5.5$, which increases to 131~GeV for a larger $M_{H^+} = 800$~GeV and $\tan\beta= 20$. These maximal values arise with zero CP-odd component in Higgs sector, as expected from our discussion above.

For values of $|X_t|/M_{\rm SUSY} \neq \sqrt{6}$, the value of $\xi_2$ may increase and the CP-odd component of the lightest neutral Higgs may be sizable. However, the parameter $\eta$ is pushed to lower values  lowering the Higgs mass.
Moreover, the existence of large $\xi_2$ or $\theta$, no matter positive or negative, will drag  $m_{H_1}^2$ further down due to mixing effects.
That's the reason why we have a anti-correlation between CP-violation and Higgs mass in the MSSM.

In Figures~\ref{MHCPodd1} and~\ref{MHCPodd2} , as before, the CP-odd component was defined to be $O_{31}$. As the mass goes down, the CP-odd component may increase but is constrained by the requirement of obtaining agreement with the measured
Higgs mass value.   Although one obtains larger values of  $m_{H_1}$   for $M_{H^+} = 800$~GeV the parabola-like upper limit on the CP-odd component of the lightest Higgs is much sharper, which implies much smaller CP-odd components in the acceptable Higgs mass range. Such a behavior is not surprising,
and reflects the decrease of the mixing angle $O_{31}$ with the charged Higgs mass, namely
\begin{equation}
O_{31} \simeq -\xi_2/M_{H^+}^2.
\label{eq:O31mh+}
\end{equation}
Rewriting the above equation in terms of the mass parameters $\mu$ and $A_t$, from Eq.~(\ref{eq:xi2}) one finds
\begin{equation}
\label{10}
O_{31} \propto -\frac{ 3 h_t^4 v^2 \sin^4\beta}{ 16 \pi^2 m_{H^+}^2}\frac{{\rm Im}(\mu A_t)}{M_{\rm SUSY}^2} \left( 1 - \frac{|X_t|^2}{6 M_{\rm SUSY}^2} \right),
\end{equation}
where we have neglected subleading terms, suppressed by $1/\tan^2\beta$ factors.

Therefore, the largest CP-violating effects that can be generated at larger $m_{H^+}$ is when $|A_t|$ and  $|\mu|$ acquire large values, while the angle arg($\mu A_f$) is fixed  to give the largest possible value of the im($\mu A_t$), but still rendering $X_t$ at acceptable values to obtain the proper Higgs mass.  For smaller values of the charged
Higgs mass, the arg($\mu A_f$) tends to be pushed to lower values,  in order to reduce the mixing effects and keep the Higgs mass in an acceptable range.

\begin{figure}[H]
    \centering
    \includegraphics[width=16cm]{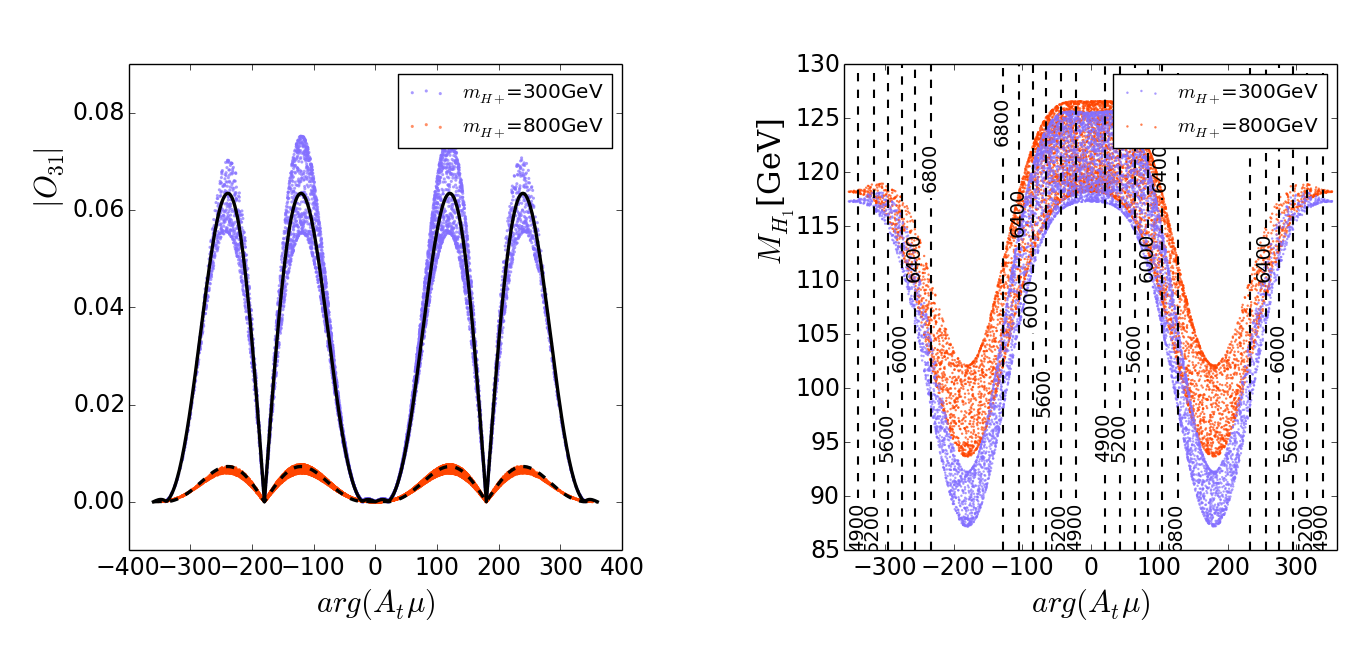}
    \caption{CP-odd component of $H_1$ and  $M_{H_1}$ as a function of the phase of  $A_t \mu$, for  $|A_t|=|\mu|=3~M_{\rm SUSY}$ and for values of other relevant parameters varied
  in the ranges given in the text. The blue and red points represent the values obtained for $M_{H^+} =$ 300~GeV and 800~GeV, respectively. The solid and dashed black lines in the left panel are the estimated value of the $H_1$ CP-Odd component by using  Eq.~(\ref{10}),
  with $h_t$ evaluated at the   $M_{H^+}$ scale, for $M_{H^+} =$ 300~GeV and 800~GeV, respectively. The dashed contour lines in the right panel represent the values of  $|X_t|$/GeV. The overall supersymmetry breaking stop mass scale $M_{\rm SUSY}$ was fixed to 2 TeV.}
    \label{fig:Fig.1}
\end{figure}

To confirm this intuition, we swept the phases of $\mu$ and $A_f$ from $-180^\circ$ to $+180^\circ$ but fixed the modulus of both $\mu$ and $A_f$ to large values, $|\mu| = |A_f| = 3 M_{\rm SUSY}$, with $M_{\rm SUSY} = 2$~TeV and $\tan\beta = 5$. The gaugino masses were fixed to $M_1 = 200$~GeV, $M_2 = 200$~GeV and $M_3 = 2.7$~TeV and the phases of three gaugino mass terms were fixed to zero.  The left panel of Figure~\ref{fig:Fig.1} shows the variation of the lightest neutral Higgs boson CP-odd component with the arg($\mu A_t$).  We see that, if the Higgs mass constraint is ignored, a maximum of the CP-odd component is obtained for phases larger than 90 degrees, actually near 120 degrees. The reason for that lies in  Eq.(\ref{10}). The dependence of $O_{13}$ on this phase is parametrized by the multiplication of two terms, Im($\mu A_t$)  and $\left(1- |X_t|^2/(6 \ M_{\rm SUSY}^2)\right)$. It is easy to show
that for the parameters chosen the maximum moves away from a phase of 90 degrees, since larger values of the
 product of these terms may be obtained by decreasing Im($\mu A_t$) but increasing the second term.  The analytical extremes for $|\mu| = |A_t| = 3 M_{\rm SUSY}$ and
 $\tan\beta = 5$ are located at values of $\phi_{A\mu} \equiv$ arg($\mu A_t$) such that $\cos \phi_{A\mu}  \simeq - 0.5$ and $\cos\phi_{A\mu} \simeq 0.94$.  This correspond to arg($\mu A_t) \simeq 120^\circ$~and~240$^\circ$ (maxima), and
 20$^\circ$ and 340$^\circ$ (minima), respectively.
   To verify this effect, we plotted Eq. ~(\ref{10}) as a function of arg($\mu A_t$) on top of the left panel of Fig.~(\ref{fig:Fig.1}) (the dashed line for $m_{H^+}=$~800~GeV and the solid line for $m_{H^+}=$~300~GeV).
 In each case, the top Yukawa coupling was chosen at the charged Higgs mass scale.  We find that Eq.~(\ref{10}) describes within a good approximation the lightest neutral Higgs CP-odd component computed by CPsuperH.

\begin{figure}[H]
    \centering
    \includegraphics[width=12cm]{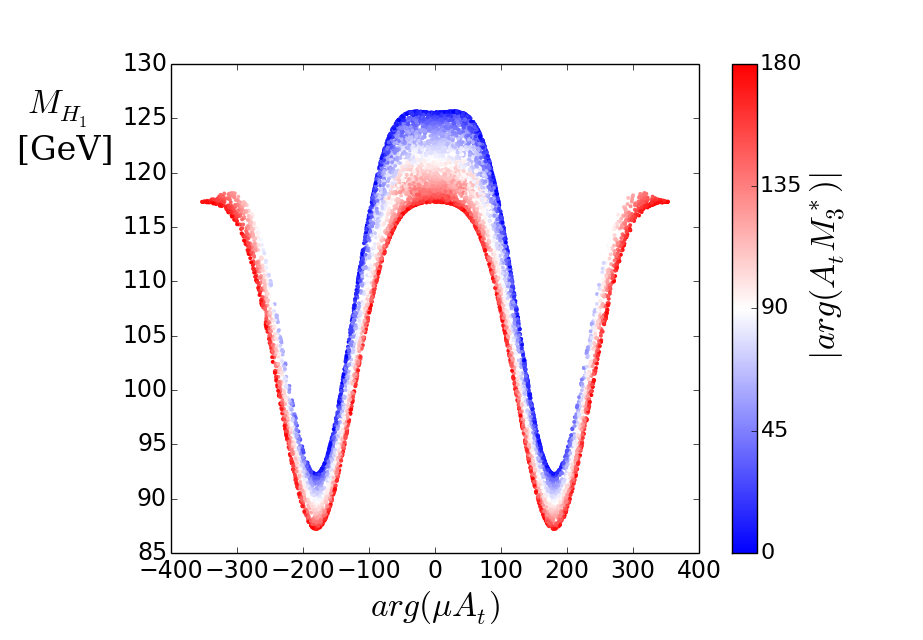}
    \caption{ Values of the Higgs mass for  $M_{H^+} =  300$~GeV, corresponding to the right panel of Fig.~\ref{fig:Fig.1}, but with the scattered points  colored according to the value of arg($A_t M_3^*$).
    The subdominant dependence of the Higgs mass on arg($A_t M_3^*$)  explains the spread of the Higgs mass values in Fig.~\ref{fig:Fig.1}. We can see an enhancement of Higgs mass when arg$(A_t M_3^*)= 0$
    and a minimum for values of arg($A_t M_3^*$) = $\pm 180^\circ$.}
    \label{fig:Fig.2}
\end{figure}

Consistency with the observed Higgs mass puts additional constraints on arg($\mu A_t$).
The right panel of Figure~\ref{fig:Fig.1} shows the strong dependence of the Higgs mass on the amplitude of $X_t$ for both $m_{H^+}=300$~GeV and $m_{H^+}=800$~GeV. Since $M_{\rm SUSY}=2~TeV$, the maximization of Higgs mass occurs close to $|X_t|=4.8$~TeV, about $2.4 \ M_{\rm SUSY}$, which is consistent with our analysis above and for $|\mu| = |A_t| = 3 M_{\rm SUSY}$ and $\tan\beta = 5$
corresponds to a phase of $\mu A_t$ close to zero. As the phase increase the CP-odd component increases, but the Higgs mass decreases. In order to keep the Higgs mass within the acceptable range, one needs $|X_t| < 6$~TeV, and should keep $|$arg($\mu A_t$)$|$ below 80 degrees, putting a bound on the possible CP-odd component of the lightest Higgs boson. This bound is about 5~percent in the particular case of $M_{H^+} = 300$~GeV.

Observe that the Higgs mass is not a single-valued function of $|X_t|$ but for each $|X_t|$ the Higgs mass values are within a broad band,  which is due to the fact that 
there are small changes in the lightest Higgs mass induced by the variation in the phase of $A_t M_3^*$, and mostly coming from threshold corrections to the top Yukawa coupling.
 An example of this variation is shown in Figure~\ref{fig:Fig.2}, where we show that indeed, besides the overall dependence on $X_t$, which is fixed by the phase of $\mu A_t$, there is a dependence on the phase of $A_t M_3^*$ leading to larger Higgs mass values for these phases equal to zero. Observe that, since
 this effect does not depend on the sign of the arg($A_t M_3^*$), in Figure~\ref{fig:Fig.2} we present the results as a function of $|$arg($A_t M_3^*$)$|$.

\section{constraints from the Higgs $H_1$ branching ratios}
\label{sec:decay}

As stressed above, a large CP-odd component of the lightest neutral Higgs may only be obtained for low values of the charged Higgs mass. Such values of
the charged Higgs mass lead in general to large mixings not only with the would-be CP-odd Higgs but also between the two would-be CP-even Higgs bosons.
Since the would-be CP-odd Higgs and the  heavier would-be CP-even Higgs have $\tan\beta$ enhanced couplings to the down fermions, in general one
expects significant deviations of the down couplings of the lightest neutral Higgs with respect to the SM one. This can be seen by writing the down-quark
couplings~\cite{Carena:2000yi}, normalized to the SM values, in the Higgs basis
\begin{eqnarray}
g^S_{H_1 dd} & = & \frac{1}{h_d+\delta h_d+\Delta h_d \tan\beta}\left\{Re(h_d+\delta h_d) \frac{-\sin\beta O_{21} + \cos\beta O_{11}}{\cos\beta} \right.
\nonumber\\
& + & \left. Re(\Delta h_d) \frac{O_{21} \cos\beta + O_{11} \sin\beta}{\cos \beta}
 -\left[Im(h_d+\delta h_d)\tan \beta-Im(\Delta h_d)\right]O_{31}\right\}
\end{eqnarray}
\begin{eqnarray}
g^P_{H_1 dd} & = & \frac{1}{h_d+\delta h_d+\Delta h_d \tan\beta}\left\{\left(Re(\Delta h_d)-Re(h_d+\delta h_d) \tan \beta\right) O_{31}\right.
\nonumber\\
& - & \left. Im(h_d+\delta h_d)\frac{-\sin\beta O_{21} + \cos\beta O_{11}}{\cos\beta}-Im(\Delta h_d)\frac{O_{21} \cos\beta + O_{11} \sin\beta}{\cos \beta}\right\},
\end{eqnarray}
where we have assumed that
\begin{equation}
h_d + \delta h_d + \Delta h_d \tan\beta = \frac{m_d \sqrt{2}}{v}
\end{equation}
is real and positive. For moderate or small values of $\tan\beta$ one can in a first approximation ignore the small radiative correction effects and, hence
\begin{eqnarray}
g^S_{H_1dd} & \simeq & O_{11} - \tan\beta \ O_{21}
\nonumber\\
g^P_{H_1dd} & \simeq & -O_{31} \tan\beta.
\label{eq:gSgP}
\end{eqnarray}
Then, as anticipated,  the corrections to the down-quark and charged lepton couplings are proportional to the non-standard components of the
lightest neutral Higgs, $O_{21}$ and $O_{31}$, but enhanced by a $\tan\beta$ factor.  Morever, while $O_{31}$ is approximately given by Eq.~(\ref{eq:O31mh+}),
\begin{equation}
O_{21} \simeq -\frac{\theta}{m_{H^+}^2} .
\label{eq:O21mh+}
\end{equation}

\begin{figure}[h!]
    \centering
    \includegraphics[width=9.5cm]{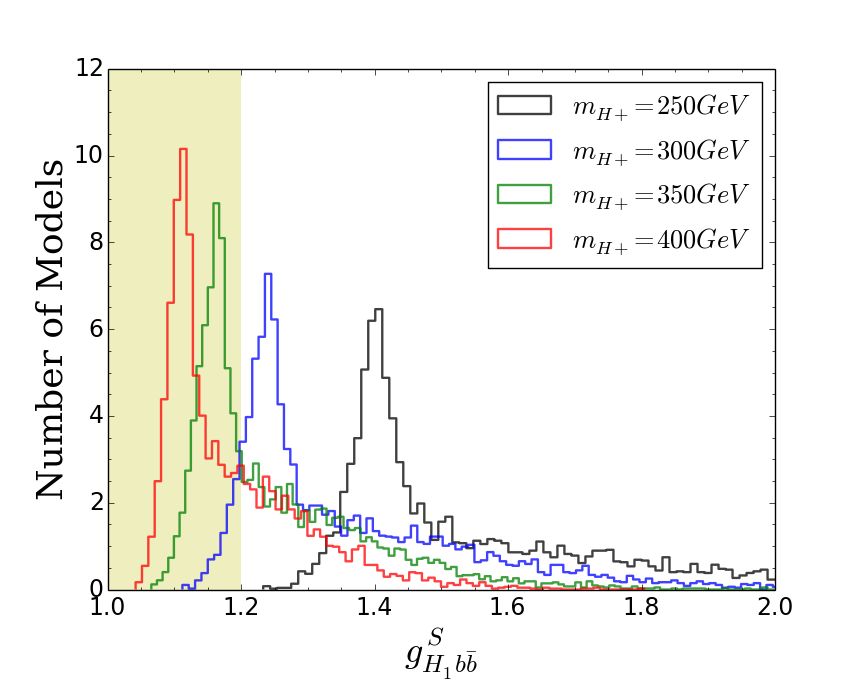}
    \caption{$g^{S}_{H_1 b\bar{b}}$ coupling for different values of $m_{H^+}$.  We have fixed $|A_f|=3M_{SUSY}$= 6~TeV; varying $|\mu|$ from 2 to 6TeV, and $\Phi_A,\Phi_{M2},\Phi_{M3},\Phi_{\mu}$ from $-180^{\circ}$ to $180^{\circ}$.}
    \label{hbb}
\end{figure}

As we can see from Fig.\ref{hbb},
the scalar coupling of the lightest Higgs boson, $g^{S}_{H_1 b\bar{b}}$, normalized to its SM value,  can grow significantly when $m_{H^+}$ is pulled down. Large deviations, however, are in tension with current experimental measurements~\cite{Khachatryan:2014jba},\cite{Aad:2014eha},\cite{Aad:2014eva}
that show a good agreement of the Higgs production rates with the SM predictions.

Since we are considering the possibility of sizable values of $\xi_2$ (the CP-odd component), the deviations from SM Higgs branching ratios may be minimized if $\theta$, which controls the mixing between two CP-even components, is kept small.  Small values of $\theta$ correspond to the condition of alignment in the case of CP-conservation~\cite{Carena:2014nza},\cite{Gunion:2002zf},\cite{Carena:2013ooa} and can be achieved for moderate values of $\tan\beta \simeq {\cal O}(10)$ if $|\mu|/M_{\rm SUSY}$ and $|A_t|/M_{\rm SUSY}$ become sizable. However, as we shall see, for alignment to happen with $|A_t|$ and $|\mu|$ smaller than 3~$M_{\rm SUSY}$, Re($A_t \mu$) must be maximized. Since maximal values of this quantity are obtained for small values of Im($A_t \mu$) controlling the CP-odd component of the lightest Higgs,  there must be some correlation between the CP-odd component of $H_1$ and the deviation of the $H_1$ down quark couplings with respect to the SM-ones.  We can obtain an
analytical understanding of this correlation by approximating  the mass of the lightest Higgs by
\begin{equation}
m_{H_1}^2 \simeq M_Z^2 \cos^22\beta + \eta,
\end{equation}
with $\eta$ given in Eq.~(\ref{eq:eta}). This is what   happens for small or moderate mixing in the neutral Higgs sector.  One can now rewrite Eqs.~(\ref{eq:theta}) and (\ref{eq:xi2}) as
\begin{equation}
\label{eq:theta1}
\theta  = \frac{1}{\tan\beta} \left[- M_Z^2 \cos 2\beta    +  m_{H_1}^2 +  \frac{3 h_t^4 v^2 \sin^4\beta}{16 \pi^2 }
{\rm Re} \left( \frac{X_t (Y_t^*- X_t^*) }{M_{\rm SUSY}^2} \left(1  - \frac{|X_t|^2}{6 M_{\rm SUSY}^2} \right) \right) \right] ,
\end{equation}
\begin{equation}
\label{eq:xi21}
\xi_2 =  \frac{1}{\tan\beta}
 \frac{ 3 h_t^4 v^2 \sin^4\beta}{16 \pi^2 } {\rm Im} \left(
\frac{ X_t (Y_t^*-X_t^*)}{M_{\rm SUSY}^2}\left(1  - \frac{|X_t|^2}{6 M_{\rm SUSY}^2}  \right) \right).
\end{equation}
Since for moderate or large values of $\tan\beta$,  $X_t \simeq A_t$, $Y_t^* - X_t^* \simeq \mu \tan\beta$ and $\cos2\beta \simeq -1$, one can see that the parameter $\theta$ can only be reduced if the real part of a loop suppressed quantity proportional to Re($A_t \mu$)  is of order of $m_{H_1}^2 + M_Z^2$.
This loop suppressed quantity is the same one whose imaginary part controls the CP-odd component.   Hence, when $\xi_2$ becomes sizable, quite generally $\theta$ cannot be suppressed and becomes also sizable. Therefore, from Eqs.~(\ref{eq:O31mh+}), (\ref{eq:O21mh+}) and (\ref{eq:gSgP}), we conclude that a significant CP-odd component in general leads to large deviations of the bottom coupling to $H_1$ with respect to the SM value.

The deviation of the $H_1$ couplings to the gauge bosons with respect to the SM ones depend on $O_{21}^2$ and $O_{31}^2$, which are in general small quantities, much smaller than the parameters controlling the deviation of the bottom and tau  couplings.
It is then expected that for moderate or large values of $\tan\beta$ the variation in the BR($H_1 \to VV$), with $V = W, Z, \gamma$, is mainly governed by the variation of the bottom quark coupling to $H_1$.
The deviation in $H_1$ down quark coupling with respect to the SM can then be inferred by the observed branching ratios of the lightest neutral Higgs to gauge bosons, namely $H\to WW^*$, $H\to ZZ^*$,$H\to \gamma\gamma$, which have been measured at the LHC up to rather high confidence
level~\cite{Khachatryan:2014jba},\cite{Aad:2014eha},\cite{Aad:2014eva}.

\begin{figure}[H]
    \centering
    \includegraphics[width=16cm]{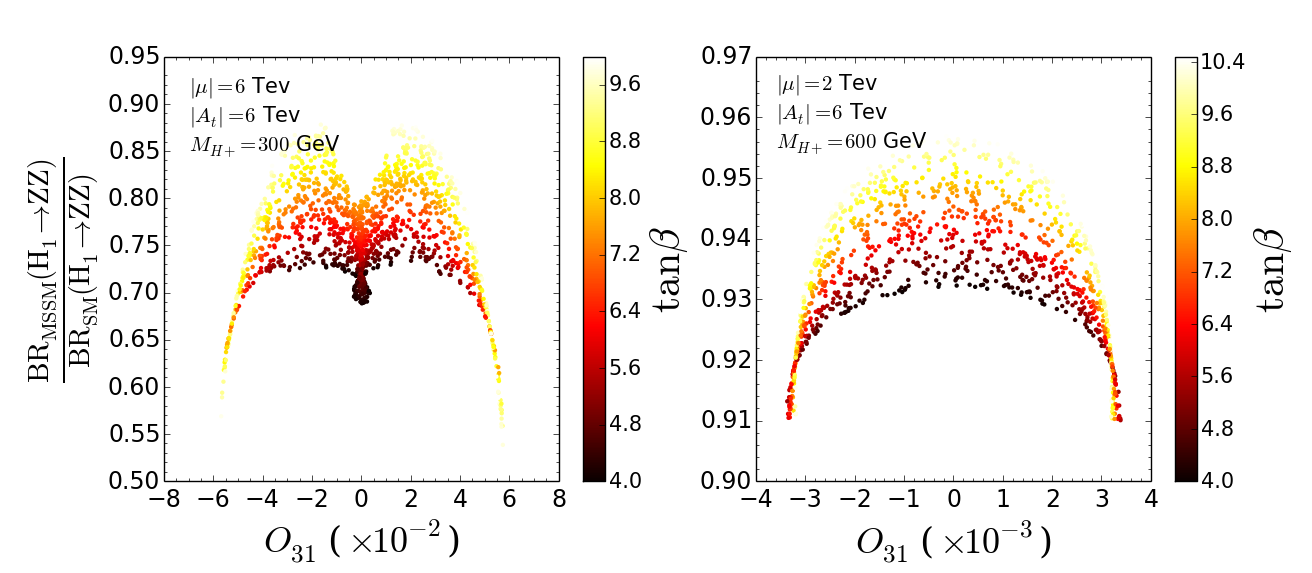}
    \caption{Correlation between the CP-odd component of $H_1$ and the $H_1$ decay branching ratio in the ZZ channel. The left panel shows the case when $m_{H^+}=300$~GeV, $|\mu|=3 M_{\rm SUSY}$=6~TeV, while the right panel corresponds to $m_{H^+}=600$~GeV and $|\mu|=M_{\rm SUSY}$=2~TeV. In both scans, we have varied the  phase of $\mu$ and the value $\tan\beta$, while  the rest of the relevant parameters were fixed to the values shown on the plot. All points shown here satisfy our $M_{H_1}$ constraint(122.5-128.5~GeV). The different colors represent different values of $\tan\beta$.  }
    \label{fig300GeV}
\end{figure}

We calculated the $H\to ZZ^*$ branching ratio in the MSSM using CPSuperH2.3 and also its value predicted by the SM for the same Higgs mass. We plotted the correlation between
the CP-odd component of $H_1$ and its decay branching ratio into $Z$ gauge bosons.
In the left panel of Fig.~\ref{fig300GeV} we show the dependence of these quantities on the variables $\tan\beta$ and $\Phi_\mu$. $\tan\beta$ is varied from 4.0 to 10.0 and $\Phi_\mu$ from $-180^\circ$ to $180^\circ$. Other parameters are chosen to maximize the Higgs mass i.e. arg($A_t M_{\tilde{g}}^*) \simeq 0$, (in this particular
example the choice of  $\Phi_A=-177.9^{\circ}$ and  $\Phi_{M_{\tilde{g}}}=173.9^\circ$ came from a scan of parameters to be presented below). Seen from this plot, the variation of $\tan\beta$ determines the shape of the arch, while $\Phi_\mu$ explains the spreading along the axis of the CP-odd component.
A correlation between the lightest Higgs boson CP-odd component  and  its branching ratio into gauge bosons  is thus observed for each independent $\tan\beta$, more specifically, the larger CP-odd component is chosen, the lower becomes the branching ratios, i.e. the more deviated from the SM values. The requirement that these branching ratios do not deviate by more than 30\% of the SM values sets a constraint for the CP-odd component of $H_1$, which according to
Fig.~\ref{fig300GeV} is tightly below 5\% for $M_{H^+} = 300$~GeV.

For comparison, in the right panel of Fig.~\ref{fig300GeV} we present the results for smaller values of $|\mu|$ and larger values of the
charged Higgs mass, namely $|\mu| = M_{\rm SUSY} = 2$~TeV and $m_{H^+} = 600$~GeV.  The value of the stop mixing parameter
was kept at $|A_t| = 3$~$M_{\rm SUSY}$.
The values of the CP-odd component
are reduced by an order of magnitude with respect to the case described in the left panel, as it is expected from the fact that $O_{31}$ is
proportional to $|\mu|/m_{H^+}^{2}$.  There is an additional small reduction, associated with the fact that
for  this value of $|\mu|$ the possible range of $\Phi_{A_t \mu}$ required to obtain values of   $|X_t|$ consistent with the $m_{H_1}$ constraints
is smaller than in the previous case.
On the other hand, the branching ratio $BR(H_1 \to ZZ)$ becomes closer to the SM value.  Due to the correlation between $O_{13}$ and the
deviation of the $H_1$ decay branching ratios with respect to the SM ones discussed above,  if in the future LHC constrains the $H_1$ decay
branching ratios to be closer to the SM ones, this will lead to further constraints on the possible CP-odd component of $H_1$.
In the following, we shall concentrate on finding the maximal value of the CP-violating phase consistent with present constraints.

Under the above considerations,  a careful scan of the whole parameter space was conducted to find the maximum  CP-odd component of $H_1$.
In order to maximize it, we chose as low values of $m_{H^+}$ as possible and for each fixed $m_{H^+}$ we scan $\tan\beta$ within the area not excluded by heavy Higgs boson searches. Since all what matters are relative phases, and the CP-violating effects are maximized for large values of $|\mu A_t|$, we fixed $M_{Q}=M_{U}=M_{D}=M_{\rm SUSY}=2$~TeV, $|\mu|=|A_t|=3~M_{\rm SUSY}$, $M_{1}=0.2$~TeV, $M_{2}=0.2$~TeV. All five varied parameters can be found in the table. The maximal CP-odd component for each scan is listed in Table \ref{table:1} and \ref{table:2}.

In Table \ref{table:1}, we show the results without including the constraints from the $H_1$ branching ratios. For all values of $m_{H^+}$, the larger CP-odd component of $H_1$ is obtained when the lightest Higgs mass reached the lower bound we have set, i.e. 122.5~$GeV$, due to the tension between a large CP-odd component  and a large enough $H_1$ mass we have proved before. As $m_{H^+}$ goes up, we see that $\Phi_{\mu A_{f}}$ is moving closer to 120$^{\circ}$ (or 240$^{\circ}$).  That's because $m_{H^+}$ is bringing up the mass of the lightest Higgs and allowing more fluctuation range in $\Phi_{\mu A_{f}}$. However the value of the $H_1$ CP-odd component  gets lower because the suppression coming for a larger $m_{H^+}$ greatly compensates the  impact of a larger phase $\Phi_{\mu A_{f}}$.
\begin{table}[H]
    \caption{Maximum CP-odd(only mass constraint)}
    \centering
    \begin{tabular}{c|ccccc|cccc}
    \hline\hline
    $m_{H^+}$(fixed) &$\tan\beta$ & $\Phi_{A_{f}}$ & $\Phi_{\mu}$ & $|M_{\tilde{g}}|$
     & $\Phi_{M_{\tilde{g}}}$ & $\Phi_{\mu A_{f}}$ & CP-odd & Mass &   $\frac{BR_{\rm MSSM}\left(H_1\to ZZ\right)}{BR_{\rm SM}\left(H_1\to ZZ\right)}$\\
    \hline
    250 & 8.0 & 158.2$^{\circ}$ & 114.0$^{\circ}$ & 3000.0 & 134.0$^{\circ}$ & 272.2$^{\circ}$ &8.87\%  & 122.6 & 0.469 \\
    300 & 9.2 & 98.8$^{\circ}$  & 2.67$^{\circ}$  & 3000.0 & 108.6$^{\circ}$ & 101.5$^{\circ}$ &5.72\%  & 122.6 & 0.555\\
    350 & 9.0 & 138.2$^{\circ}$ & 115.9$^{\circ}$ & 3000.0 & 129.5$^{\circ}$ & 254.1$^{\circ}$ &3.87\%  & 122.5 & 0.656 \\
    400 & 8.7 & 66.6$^{\circ}$  & 39.3$^{\circ}$  & 3000.0 & 76.5$^{\circ}$  & 106.0$^{\circ}$ & 2.81\% & 122.6 & 0.739\\
    \hline
    \end{tabular}
    \label{table:1} 
  \end{table}

In Table \ref{table:2}, we added the constraint on the $H_1$ decay branching ratios, which lead to somewhat smaller CP-odd components for each fixed $m_{H^+}$. For $m_{H^+}=$ 250~$GeV$ and 300~$GeV$, we see the branching ratio bound dominates the selection of the right Higgs mass and for the maximum $H_1$ CP-odd components, the Higgs mass tends to be pushed away from its theoretical lower bound. For $m_{H^+}=$ 350~$GeV$ and 400~$GeV$, instead, the Higgs mass is still the main constraint for CP violation.  The maximum value of the CP-odd component appears for charged Higgs masses of about 300~$GeV$ given both constraints. The trend in $\Phi_{\mu A_{f}}$ is the same as that in table \ref{table:1}.

\begin{table}[H]
    \caption{Maximum CP-odd (mass + Boson coupling constraints)}
    \centering
    \begin{tabular}{c|ccccc|cccc}
    \hline\hline
     $m_{H^+}$(fixed) &$\tan\beta$ & $\Phi_{A_{f}}$ & $\Phi_{\mu}$ & $|M_{\tilde{g}}|$
     & $\Phi_{M_{\tilde{g}}}$ & $\Phi_{\mu A_{f}}$ & CP-odd & Mass &   $\frac{BR_{\rm MSSM}\left(H_1\to ZZ\right)}{BR_{\rm SM}\left(H_1\to ZZ\right)}$\\
    \hline
    250 & 8.3 & 18.3$^{\circ}$ & -78.1$^{\circ}$ & 3000.0 & 17.7$^{\circ}$ &300.2$^{\circ}$ &4.83\% & 126.6 & 0.703\\
    300 & 9.5 & -177.9$^{\circ}$ & -94.0$^{\circ}$ & 3000.0 & 173.9$^{\circ}$ &88.1$^{\circ}$ &5.01\% & 124.4 & 0.701\\
    350 & 7.8 & -44.3$^{\circ}$ & -53.8$^{\circ}$ & 3000.0 & -52.1$^{\circ}$ &261.9$^{\circ}$ &3.80\% & 122.6 & 0.709\\
    400 & 8.7 & 66.6$^{\circ}$ & 39.3$^{\circ}$ & 3000.0 & 76.5$^{\circ}$ &106.0$^{\circ}$ &2.81\% & 122.6 & 0.739\\
    \hline
    \end{tabular}
    \label{table:2} 
  \end{table}

\section{Constraints on Higgs CP violation from Electric Dipole Moment experiments}
\label{sec:edm}

In addition to the collider results on the high-energy end, low-energy experiments, especially the Electric Dipole Moment (EDM) measurement with extremely high precision, impose strong constraints on the CP violation in the Higgs sector (see for instance Refs.~\cite{Brod:2013cka},\cite{Bian:2014zka}). In this section we shall explore the constraints on the possible CP violation in the MSSM Higgs sector given the present bounds on the electron EDM (eEDM),  the neutron EDM and the Mercury EDM, namely~\cite{Baker:2006ts}--\cite{Wang:2014hba}.

\begin{equation}
 \begin{split}
  \left|\frac{d_n}{e}\right| < 2.9\times10^{-26} {\rm cm}   ~(~95\%~{\rm confidence~level})\\
  \left|\frac{d_{Hg}}{e}\right| < 3.1\times10^{-29} {\rm cm}   ~(~95\%~{\rm confidence~level})\\
  \left|\frac{d_e}{e}\right| < 8.7\times10^{-29} {\rm cm}   ~(~90\%~{\rm confidence~level})
 \end{split}
\end{equation}

Theoretical calculations show that the primary contributions to EDM come from both one-loop and two-loop diagrams~\cite{Ellis:2008zy}. The dominant two-loop contributions come from the so-called Bar-Zee type diagrams~\cite{Barr:1990vd},\cite{Stockinger:2006zn}(there are other two-loop contributions~\cite{Yamanaka:2012ia}, not included in CPsuperH, which become subdominant in the regime we are working on). The most important two-loop term comes from top-quark, chargino and top-squark loop effects.  The large Yukawa coupling of the 3rd generation particles induces a large two-loop amplitude comparable to the one-loop contribution. The dominant two-loop electric dipole moment contributions are proportional to the same CP-violating phases which governs the CP violating strength in the Higgs sector, contrary to the one-loop contributions which are governed by CP-violating phases associated to particles  that couple only weakly to the Higgs fields.
In other words, large CP violation effects in the Higgs sector are likely to be associated with large two-loop EDM contributions, beyond the experimentally observed limits and could be therefore constrained by EDM experiments.

Therefore, to allow for large CP-violation effects in the Higgs sector we may need to resort to cancellations between one-loop and two-loop EDM contributions. The main one-loop contributions are from those diagrams involving loops of charginos, neutralinos and gluinos with first and second generation sfermions~\cite{Abel:2001vy},\cite{Ibrahim:2007fb}.  Therefore, the amplitudes of one-loop diagrams are in part determined by the mixing in the mass eigenstates of charginos and neutralinos, which is associated with the values of $\mu, M_1, M_2, tan\beta$, and in particular the phases ${\rm arg}(\mu M_i)$,
which also affect the two loop chargino and neutralino contributions. The one-loop contributions
decrease for heavier first and second generation squarks and sleptons. As we said before, we shall characterize the ratio of the first and second to the third generation sfermion masses by a hierarchy factor $\rho$, which is an input parameter in the CPsuperH code. . 

\begin{figure}[H]
    \centering
    \includegraphics[width=9.5cm]{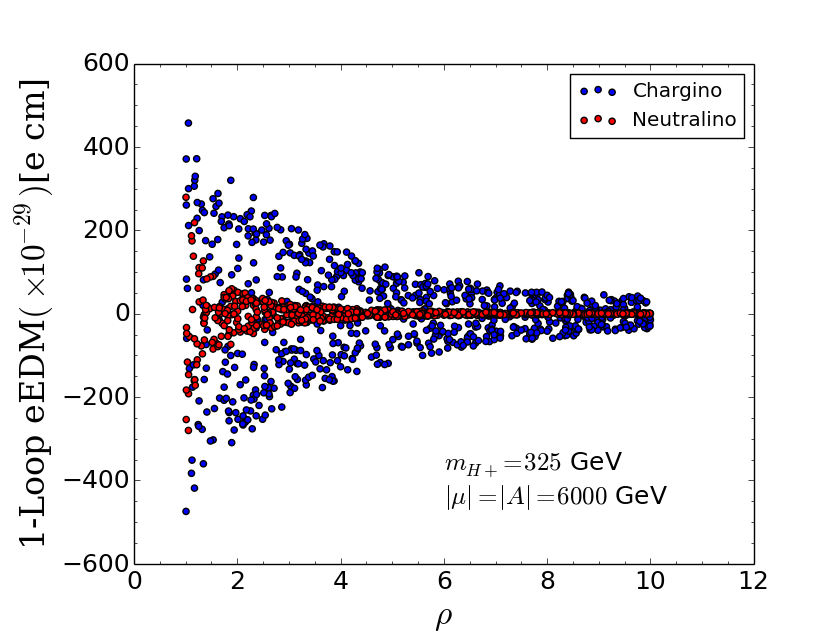}
    \caption{One-loop contribution to the electron EDM.  All the points shown in this plot lead to a value of $M_{H_1}$ compatible with the observed Higgs mass.  The relevant parameters are fixed as follows :
    $m_{H^+}$ is fixed at 325~GeV,
    $|\mu|=|A|=3 M_{\rm SUSY}$=6~TeV, $\tan\beta$
    is varied  from 4 to 9, $\Phi_{\mu},\Phi_{A},\Phi_{M_2},\Phi_{M_3}$ are varied from -180 to 180 and  $|M_3|$ from 1.5~TeV to 3~TeV.  }
    \label{edmchneu}
\end{figure}

In Figure~\ref{edmchneu} we display the one-loop contribution to the electron EDM.
From Figure~\ref{edmchneu}, we find that, as expected, both the one-loop chargino and neutralino contributions to the electron EDM decrease as we raise $\rho$. Up to $\rho$. The maximum chargino contribution remains higher than the acceptable eEDM limit ($8.7\times10^{-29} {\rm cm}$) up to values of $\rho = {\cal{O}}(10)$. Another feature seen from this plot is that the amplitude of chargino-loop diagrams is pronouncedly larger than that of neutralino-mediated ones, differing by an order of magnitude. Thus, unless the phases are highly fine tuned, it is very difficult for EDM to cancel within one-loop level diagrams.

\begin{figure}[H]
    \centering
    \includegraphics[width=17cm]{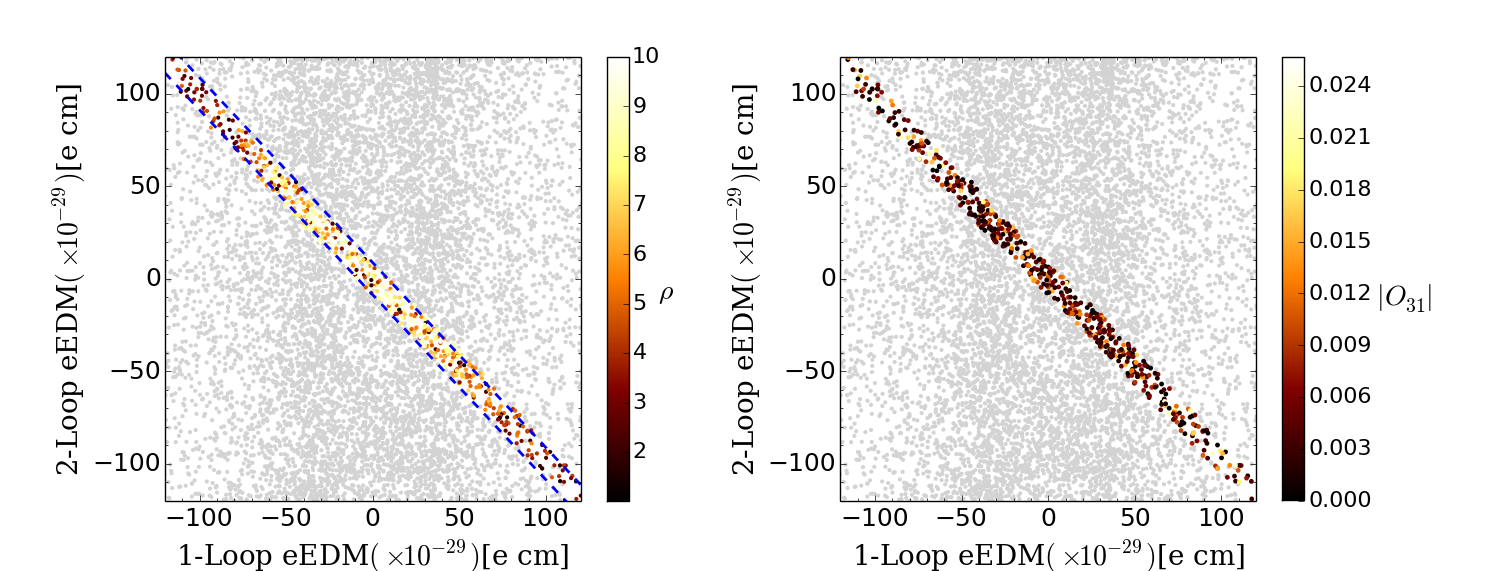}
    \caption{Correlation between 1-loop and 2-loop Contributions to the electron EDM('eEDM' in the axis labels stands for electron EDM). All the points shown give appropriate $H_1$ mass values, among which the colored ones satisfy the electron EDM bound of $8.7\times10^{-29} {\rm e~cm}$. This is the same scan as in Fig.~\ref{edmchneu}. The colors in the left panel represent the values of  $\rho$ and in the right panel the $H_1$  CP-Odd component. This plot illustrates that the eEDM constraint can be avoided by cancellation between the 1-loop and 2-loop contributions.}
\label{onevstwoedm}
\end{figure}

The left panel of Fig.~\ref{onevstwoedm} shows the correlation between the one and two-loop contributions to the electron EDM for parameters which survive the current bounds on this quantity(eEDM$<8.7\times10^{-29} {\rm e~cm}$).  Points in this figure are  colored according to  the value of the hierarchy factor $\rho$.  In the right panel we show the same correlation  but points are colored according to the size of the CP-odd component of the lightest neutral Higgs boson.

\begin{figure}[H]
    \centering
    \includegraphics[width=9cm]{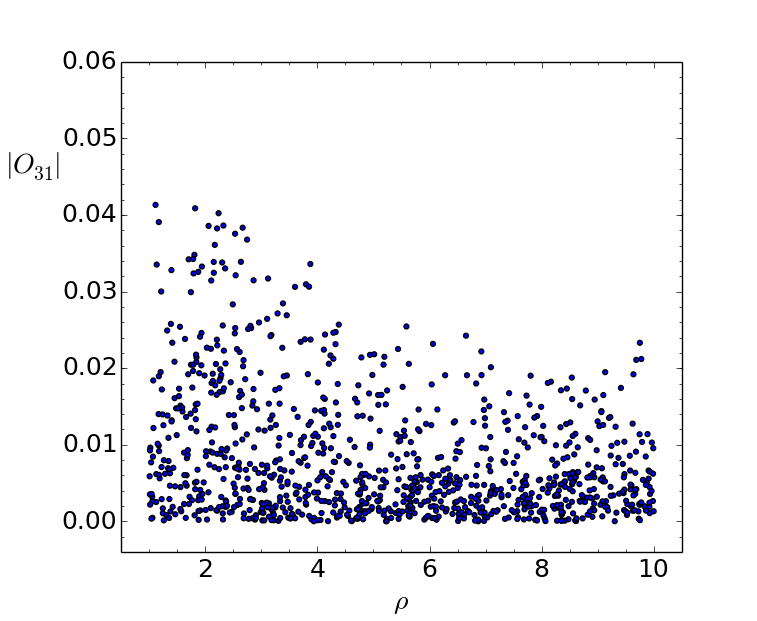}
    \caption{The $H_1$ CP-odd component vs the hierarchy factor $\rho$, using the same scan as  in Fig.~\ref{edmchneu}. All the points shown in this plot give appropriate $H_1$ mass values and  satisfy the eEDM bound. }
    \label{edmvsrho}
\end{figure}

We find that most of the allowed points lie closely around a straight line across the origin point with slope $-1$ which indicates that an approximately exact cancellation occurs between one-loop and two-loop contributions to the electron EDM.
Figure~\ref{edmvsrho} shows the correlation between the CP-odd component of $H_1$ and the hierarchy parameter $\rho$.
As shown in the right panel of Fig.~\ref{onevstwoedm} and Fig.~\ref{edmvsrho}, larger CPV coexists with larger two-loop (or one-loop) EDM components and, for third generation squark masses of the order of one TeV,  appears around a $\rho=2$ peak, where the one-loop contributions are sizable and cancellations between one and two-loop contributions are significant. Therefore the possibility of a pronounced CP-violating effect in the Higgs sector relies on significant cancellations between one-loop and two-loop EDM contributions.

In order to explore the maximum allowed  CP-odd component of $H_1$ given currently measured EDMs, we use CPSuperH2.3 to  scan over all relevant variables, choosing low values of the charged Higgs mass and large values of the stop mixing.  More specifically, we chose $M_{\rm SUSY}=2$~TeV (including all squark and slepton masses) and $|\mu|=|A_f|=3 \ M_{\rm SUSY}$.  The electroweak gaugino masses  values were fix at  $|M_1|=|M_2|=200$~GeV, $\Phi_{M_{1}}=0$ (since only the relative phases matter), and the charged Higgs mass was fixed at $m_{H^+}=300$~GeV so that we can get sizable CP violation and also keep $BR(H \to VV)$ within an acceptable range at the same time. The value of $\tan\beta$ was varied from 5.5 to 9.5 (consistent with the current experimental bounds), the hierarchy factor $\rho$ was varied  between 1 and 10, while $|M_3|$ was varied from 1.5~TeV to 3~TeV. The phases of the mass parameters $\Phi_{A_f}$, $\Phi_{\mu}$, $\Phi_{M_3}$, $\Phi_{M_2}$ were varied from $-180^{\circ}$ to $+180^{\circ}$.  To fight against the high elimination rate associated with the experimental constraints and the huge complexity in computing EDMs, we implemented a gradient descent method in the 3D subspace spanned by parameters $\Phi_{M_2}$, $\Phi_{\mu}$, and $\rho$ to bring the three EDM values into acceptable ranges. The descending process was fast with proper steps and iteration algorithm. Finally we found 4200 points passing all constraints, with a maximum CP-odd component of $H_1$ to be 3.07\%, which is consistent with our observations above.

To exemplify the values of the parameters leading to  relevant $O_{31}$, in Table~\ref{table:nonlin} we show some of the points with maximal $H_1$ CP-odd component, the parameters for which they are obtained, as well as the relevant parameters in the Higgs sector.
\begin{table}[H]
    \centering
    \caption{Maximum CP odd component points after taking EDM constraints into account.The values of the stop and Higgsino mass parameter were fixed to $|A_t| = |\mu| = 3 M_{\rm SUSY} = 6~TeV$.  The other relevant parameters were varied in the range explained in the text.}
    \begin{tabular}{c|ccccccc|ccc}
    \hline\hline
    $No.$ & $\tan\beta$ & $\Phi_{\mu}$ & $\Phi_A$ & $|M_{\tilde{g}}|$
     & $\Phi_{M_{\tilde{g}}}$ & $\rho$ & $\Phi_{M2}$ & $m_{h}$ & $\frac{BR_{\rm MSSM}\left(H_1\to ZZ\right)}{BR_{\rm SM}\left(H_1\to ZZ\right)}$ & CP-odd component\\
    \hline
    1 & 9.5 & $-45.7^{\circ}$ & $-18.5^{\circ}$ & 2300 & $54.2^{\circ}$ & 9.17 & $11.3^{\circ}$ & 122.7 & 0.782 & $3.00\%$\\
    2 & 9.0 & $-34.1^{\circ}$ & $-31.6^{\circ}$ & 3000 & $39.0^{\circ}$ & 3.86 & $10.2^{\circ}$ & 122.7 & 0.776 & $3.07\%$\\
    3 & 8.9 & $-2.8^{\circ}$ & $-62.7^{\circ}$ & 3000 & $9.9^{\circ}$ & 5.26 & $-18.9^{\circ}$ & 122.5 & 0.774 & $3.04\%$\\
    4 & 8.5 & $23.3^{\circ}$ & $-88.0^{\circ}$ & 3000 & $-17.0^{\circ}$ & 3.44 & $-39.9^{\circ}$ & 122.6 & 0.772 & $2.96\%$\\
    5 & 8.6 & $177.4^{\circ}$ & $-121.3^{\circ}$ & 2750 & $172.7^{\circ}$ & 8.53 & $-149.4^{\circ}$ & 123.8 & 0.796 & $2.41\%$\\
    \hline
    \end{tabular}
    \label{table:nonlin} 
\end{table}

Observe that these five different examples have similar characteristics :  The values of arg$(\mu M_{\tilde g}) \simlt 10^\circ$, as expected in order to cancel the large one-loop contribution to the neutron EDM, induced by the gluino loops. Moreover, the value of arg$(\mu M_2)$  is within 30$^\circ$ of 0 or 180$^\circ$.  The value of arg$(\mu A_t) \simeq 65^\circ$, being sizable and of similar order in all examples,  is necessary to obtain a sizable CP-odd component of $H_1$ without inducing a large negative correction  to its mass or  to the branching ratio of its decay into vector bosons.  As is shown in the table~\ref{table:nonlin} the maximal CP-odd component is now again associated with the minimal allowed values of the Higgs mass. This may be understood from the fact that, as shown in Fig.~\ref{fig:Fig.2}, the largest values of $m_{H_1}$ are associated with values of arg$(A_t M_3^*) = 0$.  However, since the electric dipole
moment constraints  lead to arg$(A_t M_3^*) \simeq {\rm arg}(A_t \mu)$, a large CP-odd component of $H_1$ leads to values of the Higgs mass away from its maximal value. Hence, the Higgs mass combined with the constraints on electric dipole moments puts an additional bound on the possible values of the $H_1$ CP-odd component.

We want to stress  that $|M_1|$ and $|M_2|$ are not determinant factors in the determination of the maximal $H_1$ CP-odd components. We changed $|M_1|=|M_2|$ to be 1~TeV but kept $|\mu|=|A_f|=3 \ M_{\rm SUSY}$, and got the maximum CP-odd to be 2.91\%, not much different from the previous 3.07\%. We also
checked the maximal CP-violation in the CPX scenario in which $|\mu|=4 \ M_{\rm SUSY}$, $|A_f|=2 \ M_{\rm SUSY}$, $|M_1|=|M_2|=1$~TeV, $M_{3}=3$~TeV, while the three trilinear coupling phases $\Phi_{A_{t,b,\tau}}$ are independent. We did the scan for this scenario and found that it gave a smaller CP-odd component of about 2\%. This effect comes mostly from the change of $|\mu|$ and $|A_f|$, which can be easily seen from Eq. \ref{10}. This agrees with the numerical results of a recent paper~\cite{Arbey:2014msa} focusing on the CP Violation in the heavy Higgs sector of the MSSM.

\begin{figure}[H]
    \centering
    \includegraphics[width=9cm]{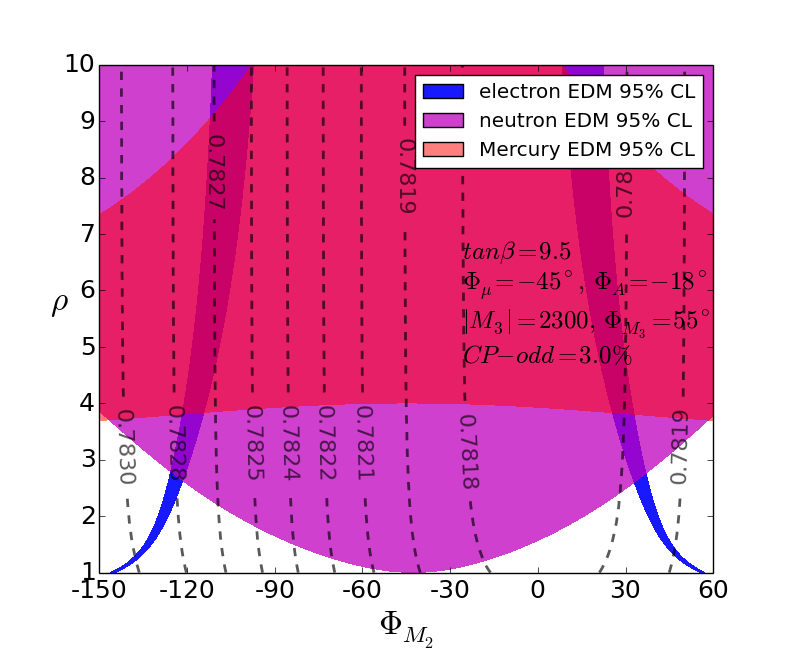}
    \caption{EDM constraints in the  $\Phi_{M_2}$--$\rho$ plane. The allowed regions for the three kinds of EDMs are drawn in different colours on this patch of the 2D parameter plane.  The other relevant parameters were chosen to be the same as in parameter set 1 in Table~\ref{table:nonlin}, i.e. $\tan\beta = 9.5$,$\Phi_{\mu} = -45^{\circ}$, $\Phi_A = -18^{\circ}$, $M_3 = 2300$~GeV, $\Phi_{M_3} = 55^{\circ}$. 
   The dashed lines show countors of the ratio $BR_{\rm MSSM}\left(H_1\to ZZ\right)/BR_{\rm SM}\left(H_1\to ZZ\right)$, which displays a tiny fluctuation of about $0.1\%$ over the whole range.}
    \label{CPoddmax1}
\end{figure}

In general, we observe that  the cancellation of the three EDMs needs some fine tuning at level of order 10 in relevant phases. In order to illustrate the general pattern of cancellations we investigate the behavior of the three EDMs around the points of maximal $H_1$ CP-odd component found above. For instance, Fig.~\ref{CPoddmax1} shows the values of the three EDMs considered here, for points around point 1 in Table III, and varying $\rho$ and $\Phi_{M_2}$ only. The mass and the lightest neutral Higgs boson CP-odd component contour lines are not shown on these plots since they are almost constant over the whole region displayed (122.7~GeV and $3.0\%$ respectively). The dashed contour line indicates the ratio of the $BR(H \to ZZ)$ to the SM values, showing acceptable values  over this whole region of parameter space. The parameters $\rho$ and $\Phi_{M_2}$ are chosen because they have nearly nothing to do with the neutral Higgs masses and affect only weakly the  CP-violation in the Higgs sector (i.e. $O_{31}$) but they affect strongly the EDMs through one-loop diagrams. As illustrated in these plots, there seems to be no difficulty in finding some combinations of $\rho$ and $\Phi_{M2}$ to circumvent the strong EDM constraints, at least for the current bounds.  All points allowed in these examples, however, have values of $\rho \simgt 4$, implying that in this example one cannot achieve the  maximum $H_1$ CP-odd component, as we showed before, which are obtained for values of $\rho \simeq 2$.

\begin{figure}[H]
    \centering
    \includegraphics[width=12cm]{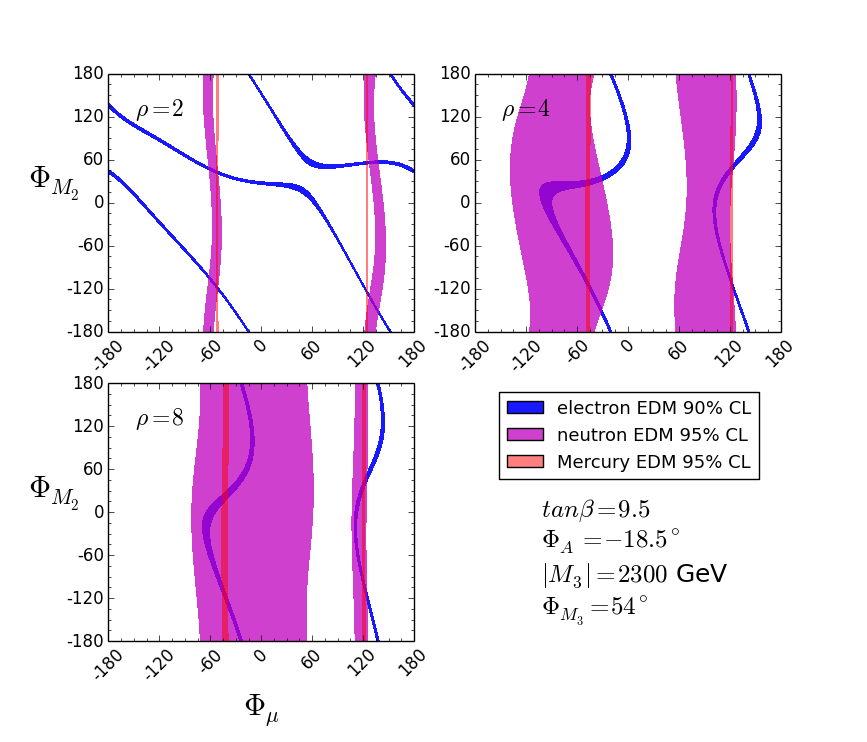}
    \caption{EDM constraints in the  $\Phi_{\mu}$ -- $\Phi_{M_2}$ plane for  different values of  $\rho$. All other relevant parameters are consistent with the ones in parameter set 1 in Table~\ref{table:nonlin} (the same as in Fig.~\ref{CPoddmax1}). 
    }
    \label{phimuvsphim2rho}
\end{figure}

Figure~\ref{phimuvsphim2rho} shows the correlation between the phases of $\mu$ and $M_2$ for the points which are consistent with the electron, neutron and mercury EDM's.
As Fig.~\ref{phimuvsphim2rho} shows, no matter what value the hierarchy factor $\rho$ takes, there is always some point where the three constraint regions overlap with each other. As $\rho$ goes up, one-loop contribution fades away and two-loop diagrams dominate since propagators of first 2 generations of squarks and sleptons only come into play in one-loop diagrams. The blue stripe allowed by eEDM measurement rotates towards constant $\Phi_{\mu}$. This phenomenon can be easily understood since $\Phi_{M_2}$ affects the mass structure of charginos and neutralinos, which control the main one-loop contributions to the eEDM. The red stripe stands for Mercury EDM, which depends only weakly on $\Phi_{M_2}$, and it grows wider as $\rho$ increases, which may be understood due to the smaller degree of cancellation between $\Phi_{\mu}$ and the gluino phase necessary to be consistent with the current experimental bounds on this quantity.

\section{Flavor Physics Constraints}
\label{sec:flavor}

The flavor physics implications of the MSSM depend very strongly on the exact
flavor structure of the soft supersymmetry breaking parameters. Small missalignments
between the squark and the quark mass matrices can induce large flavor violating
effects, without having an impact on any other observables.  Since in our work we
are considering the MSSM as a low energy effective theory, without any assumption of the
supersymmetry breaking mechanism at high energies, it is not possible to obtain
precise predictions on the flavor observables.  In order to obtain an estimate of
the flavor violating effects, we used the results of CPsuperH, which are based on
the assumption of minimal flavor violation, with additional flavor misallignments
induced by up-Yukawa effects~\cite{Barbieri:1993av}--\cite{Carena:2008ue}, which lead
to non-vanishing contributions from flavor violating couplings of the gluino with the left-handed down-quarks and scalar down-quarks.

\begin{figure}[H]
    \centering
    \includegraphics[width=10cm]{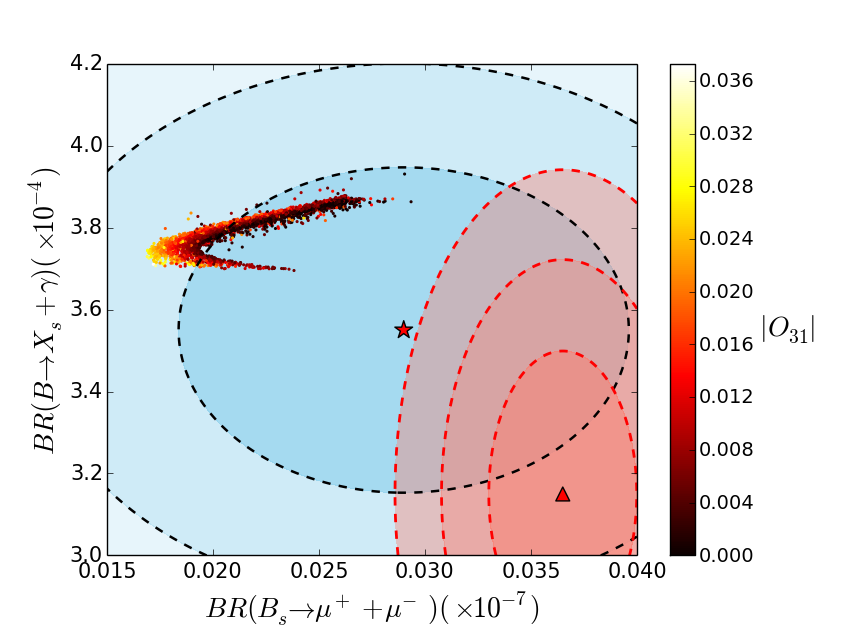}
    \caption{ The branching ratio values of the decay channels $B_s \to \mu^+ +\mu^-$ and channel $B \to X_s+\gamma$, computed in CPsuperH, are displayed for points allowed by all experimental constraints considered in this article. The points are colored by the CP-odd component of $H_1$. The red pentagram marks the current experimental values. The red triangle in the plot displays the prediction by Standard Model. The regions allowed at the
68\% and 96\% C.L. are displayed by dashed lines. }
    \label{BMeson}
\end{figure}

In general, since in the models under consideration
the squarks are heavier than about 2~TeV, $\tan\beta$ is moderate and the
charged Higgs mass is about 300~GeV, one does not expect large flavor violating effects.
These effects, however, may be enhanced by the presence of large trilinear couplings
between the Higgs and the third generation squarks. In Figure~\ref{BMeson} we show
the predictions for two relevant observables, namely the branching ratios of the
decays of $B_s \to \mu^+ \mu^-$ and $B \to X_s \gamma$.   The
current experimental values of these observables,
\begin{equation}
BR(B \to X_s+\gamma)=(3.55\pm0.24^{+0.09}_{-0.10}\pm0.03)\times 10^{-4}
\nonumber
\end{equation}
as estimated by the Heavy Flavor Averaging Group for $E_\gamma>1.6$~GeV~\cite{Barberio:2006bi},
and
\begin{equation}
BR(B_s \to \mu^+ +\mu^-)=( 2.9 \pm 0.7)\times 10^{-9}
\nonumber
\end{equation}
as recorded by LHCb and CMS analyses~\cite{CMSandLHCbCollaborations:2013pla} are
in somewhat good agreement with the SM predictions~\cite{Misiak:2006zs},\cite{Misiak:2006ab},\cite{Bobeth:2013uxa} given by
\begin{equation}
 BR(B \to X_s \gamma) = (3.15 \pm 0.23 ) \times10^{-4}
\nonumber
\end{equation}
(see Ref.~\cite{Becher:2006pu} for an alternative calculation of this rate) and
\begin{equation}
BR(B_s \to \mu^+\mu^-) = (3.65 \pm 0.23) \times10^{-9}.
\nonumber
\end{equation}
In our analysis, we performed a small rescaling of
the values of $B \to X_s \gamma$ given by CPsuperH in order to obtain the proper SM results~\cite{Misiak:2006ab} for large squark and charged Higgs masses.

In Figure~\ref{BMeson} we show with dashed lines the regions allowed at the 68\% and 96\%
confidence level (C.L.).
We see that under
the above assumptions, for the maximal CP-violating effects in the Higgs sector, the predicted values of these two observables are in good agreement with
the experimental values  and
actually this model leads to a similarly good description of these observables to  the
one obtained in the SM. Therefore, these flavor observables do not put additional constraints
on the allowed values of the CP-odd component of the lightest neutral Higgs boson.

\section{Probes of the $H_1$ CP-odd Component at the LHC.}
\label{sec:LHC}

The small CP-odd components of the lightest CP-even Higgs boson make its detection difficult.
 A variety of  observables that may lead to the determination of the $H_1$ CP-odd component have  been constructed and different experiments are proposed to measure a CP mixing directly, for example, the azimuthal angle correlations between two jets in Higgs plus two jets channel via gluon fusion~\cite{Klamke:2007cu}, the polarization correlation in the $H\to \gamma Z$ and $H \to \gamma\gamma$ channels~\cite{Korchin:2013ifa}, the angular distribution of the products in the $t\bar{t}H$ channel~\cite{Gunion:1996xu},\cite{Boudjema:2015nda}, as well as the distribution over the angle between the planes of $e^-e^+$ pairs arising from conversion in diphoton decays~\cite{Voloshin:2012tv},\cite{Bishara:2013vya}.

A promising channel, $h\to\tau^-\tau^+$, has been proposed to investigate the CP nature of the Higgs boson at the LHC~\cite{Harnik:2013aja},\cite{Berge:2014sra},\cite{Dolan:2014upa}, and becomes suitable to test CP-violation in the Higgs sector of the MSSM. In the recent  proposal, Ref.~\cite{Berge:2014sra}, the mixing angle, $\phi_\tau$, defined as:
\begin{equation}
\tan\phi_\tau=\frac{g^{P}_{h\tau\tau}}{g^{S}_{h\tau\tau}}
\label{eq:phitau}
\end{equation}
can be determined by measuring the spin correlation of the tau lepton pairs, which lead to particular differential distributions of the tau pairs in the Higgs decays.  These correlations are characterized by an angle $\phi^{*}_{CP}$, defined from the impact parameters and momenta of the charged prongs
$a^-$ and $a^+$ in the decays  $\tau^-\to a^- + X$ and $\tau^+\to a'^{+} + X$ in the $a^-a'^{+}$ zero-momentum frame. The measured  differential distribution of the Higgs boson decaying into tau-pairs with respect to $\phi^{*}_{CP}$ can be described by:

\begin{equation}
\frac{d\sigma(p p \to H_1 \to \tau\tau)}{d\phi^*_{CP}} \simeq u\cos(\phi^{*}_{CP}-2\phi_{\tau})+v
\end{equation}

The major background comes from the Drell-Yan production of $\tau$ pairs whose effects can be minimized by cuts. It is claimed that the Higgs mixing angle $\phi_\tau$ can be measured to a precision of $\Delta\phi_\tau\approx 14.3^\circ(5.1^\circ)$ at the high luminosity LHC (14~TeV) with an integrated luminosity of $500~fb^{-1} (3~ab^{-1})$ (Ref.~\cite{Harnik:2013aja}, instead, claims a sensitivity of about $11^\circ$ at 3~$ab^{-1}$).

In the Higgs basis, considering only the dominating terms, $\tan\phi_\tau$ can be approximated by
\begin{equation}
\tan \phi_\tau \simeq \frac{O_{31}  \tan\beta}{O_{11} - O_{21} \tan\beta},
\end{equation}
which leads to values of $\phi_\tau$ of order of $10^\circ$ for values of $O_{31}$ and $O_{21}$ of a few percent and $\tan\beta \simeq 10$, and grows for larger values of $\tan\beta$..
For instance, for point 1 in Table~\ref{table:nonlin}, a value of $\tan\phi_\tau = 0.236$ is obtained, corresponding to $\phi_\tau = 13^\circ$, within the reach of LHC. This is well within the
claim reach of the high luminosity LHC.

\begin{figure}[H]
    \centering
    \includegraphics[width=10cm]{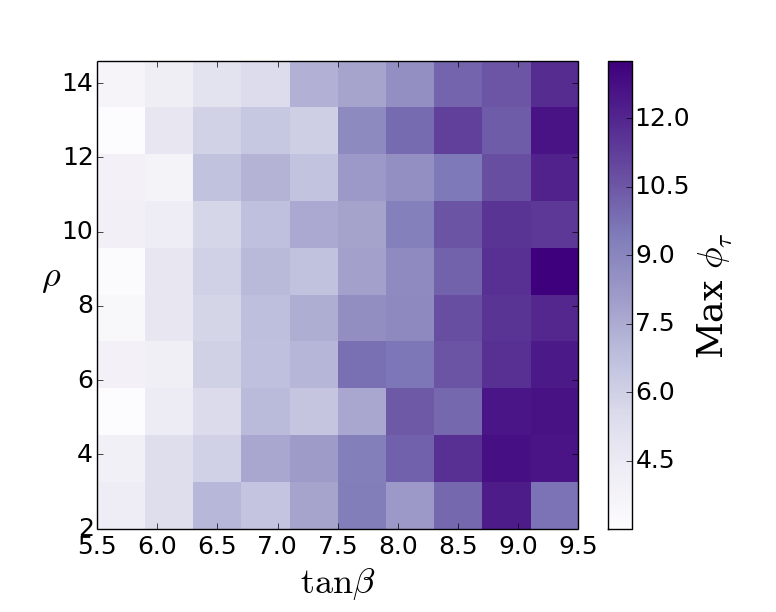}
    \caption{Maximum value of $\phi_\tau$, Eq.~(\ref{eq:phitau}), in the $\tan\beta$ - $\rho$ plane, obtained from a
a scan of the phases of all relevant parameters, $A_f$, $\mu$, $M_3$ and  $M_2$, for $m_{H^+} = 300$~GeV,  $|A_t| = | \mu | = 3 M_{\rm SUSY} = 6$~TeV. The values of
$\tan\beta$ and $\rho$ are varied within a fairly large range, and points consistent with the present experimental constraints are selected.}
    \label{MAXCPV}
\end{figure}

To get a better perception of  the power of the $h\to\tau^-\tau^+$ measurement, in  Fig.\ref{MAXCPV} we plot, for the  points we found  satisfying all current experimental constraints considered in this paper,
the maximum value of $\phi_\tau$ in the $\tan\beta-\rho$ plane. In other words, these values represent the experimental sensitivity needed  in order to start probing the CP-odd component of $H_1$ in the MSSM  for that particular parameter region.

It is then clear that if the value of $O_{31}$ is close to the maximal values consistent with current experimental constraints,  the LHC may probe this CP-violating effects in the high luminosity run. It is also clear that in order for the LHC to probe the CP-odd component of $H_1$ in the MSSM, the charged Higgs mass should be of order of the weak scale and $\tan\beta >5$.  This region of parameters will be efficiently probed by the LHC in the search for Higgs bosons decaying into $\tau$-pairs in the near future. Moreover, as stressed before a large CP-odd component of $H_1$ is in general associated with a modification of the branching ratios of $H_1$ and hence precision measurements of the $H_1$ properties will further test the region of parameter space consistent with a significant CP-odd component of $H_1$.

\section{Conclusion}
\label{sec:conclusions}

In this article, we studied the  values of the CP-odd component of the lightest neutral Higgs allowed by current experimental constraints. We
derived new analytical expressions in the Higgs basis that allow a good understanding of the parametric dependence of this component on the supersymmetry breaking parameters. We showed that the values of the stop left-right mixing parameter that maximize the lightest CP-even Higgs mass
lead to a suppression of the dominant loop contribution to the CP-odd component of the lightest Higgs boson. Since for stop masses of order of the
TeV scale, stop mixings close to the ones that maximize $m_{H_1}$ are necessary in order to obtain SM-like Higgs masses of order of the one observed
experimentally, the measured Higgs mass puts a significant constraint on the possible values of the $H_1$ CP-odd component.

Moreover, we showed that large $H_1$ CP-odd components lead necessarily to a significant increase of the width of the lightest neutral Higgs decay  into bottom quarks. Since the  width of $H_1 \to b \bar{b}$  is the dominant decay width of $H_1$, this increase leads also to a significant modification of the branching ratio of the decays of $H_1$ to gauge bosons, what leads to a further constraint into large $H_1$ CP-odd components.

Electric dipole moments put a further constraint on this possibility. Although cancellations between one-loop and two-loop contributions may lead to acceptable values of the electron EDM, which is the most precisely bounded one at this point, the strong alignment between the phases of $\mu$ and the gluino mass leads to further  restrictions on  the possible obtention of a large $H_1$ CP-odd component. At the end, we showed that the CP-odd component of $H_1$ is restricted to be smaller than about 3\%.  Furthermore, we analyzed relevant flavor physics observables and shoed that they do not set additional constraints on this $H_1$ property.

We also studied the possible experimental detection of the $H_1$ CP-odd component at the LHC. The $h \to \tau^- \tau^+$ channel presents a very efficient probe of this possibility.  The CP-odd coupling of the $\tau$ lepton to $H_1$ is proportional to the $H_1$ CP-odd component but it is enhanced by a $\tan\beta$ factor.   Due to this enhancement, we showed that, for values of the charged Higgs mass of the order of the weak scale and $\tan\beta >5$, a determination of the $H_1$ CP-odd mixing is possible at a high luminosity LHC, but only for values close to the largest allowed values of this mixing. The observation of a non-vanishing CP-odd component of $H_1$ would then put strong constraints on the parameter space of the MSSM.  Further constraints coming from precision measurement of the $H_1$ branching ratios and searches for heavy Higgs bosons may further probe the parameter space consistent with an observable CP-odd component of $H_1$ in the MSSM.

Let us emphasize in closing that the constraints on the CP-violating components of $H_1$ discussed in this paper are specific for the MSSM
and could not be generalized to more general two Higgs doublet models, where larger CP-violating effects in the Higgs sector may be present,
as has been discussed in Refs.~\cite{Klamke:2007cu}--\cite{Dolan:2014upa}.  Some of these constraints are related to the specific properties
of the radiative corrections leading to the Higgs mass generation in the MSSM and may be avoided in non-minimal supersymmetric extensions,
like the NMSSM (see for instance Ref.~\cite{Moretti:2013lya}).  Finally, while the LHC capabilities are limited, measurement of the CP-violating component of $H_1$ may be improved at lepton colliders, as was discussed in detail in Refs.~\cite{Bhupal Dev:2007is}--\cite{Ananthanarayan:2014eea}. We
plan to come back to these subjects in the near future.

\section{ACKNOWLEDGMENT}
We would like to thank the Aspen Center for Physics, which is supported by the National Science Foundation under Grant No. PHYS-1066293. Work at ANL is
supported in part by the U.S. Department of Energy under
Contract No. DE-AC02-06CH11357. Work at EFI is
supported by the U.S. Department of Energy under
Contract No. DE-FG02-13ER41958.

\newpage.

\end{document}